\documentclass[prl,aps,showpacs,twocolumn,tightenlines,superscriptaddress]{revtex4}  
\usepackage{graphics}
\usepackage{color}
\usepackage{epsf}
\def\TME{\tau^-\rightarrow\mu^-\eta}

\def\T3M{\tau \rightarrow 3\mu}
\def\EGG{\eta\rightarrow \gamma\gamma}
\def\E3P{\eta\rightarrow \pi^+\pi^-\pi^0}
\def\TT{\tau^+\tau^-}
\def\ME{\mu\eta}
\def\MM{\mu^+\mu^-}
\def\GG{\gamma\gamma}

\def\Minv{M_{\mu\eta}}
\def\DelE{\Delta E}

\begin{document}

\title{ Search for the Lepton-Flavor-Violating Decay
          $\tau^-\rightarrow\mu^-\eta$ at Belle}

\affiliation{Aomori University, Aomori}
\affiliation{Budker Institute of Nuclear Physics, Novosibirsk}
\affiliation{Chiba University, Chiba}
\affiliation{Chuo University, Tokyo}
\affiliation{University of Cincinnati, Cincinnati, Ohio 45221}
\affiliation{University of Frankfurt, Frankfurt}
\affiliation{Gyeongsang National University, Chinju}
\affiliation{University of Hawaii, Honolulu, Hawaii 96822}
\affiliation{High Energy Accelerator Research Organization (KEK), Tsukuba}
\affiliation{Hiroshima Institute of Technology, Hiroshima}
\affiliation{Institute of High Energy Physics, Chinese Academy of Sciences, Beijing}
\affiliation{Institute of High Energy Physics, Vienna}
\affiliation{Institute for Theoretical and Experimental Physics, Moscow}
\affiliation{J. Stefan Institute, Ljubljana}
\affiliation{Kanagawa University, Yokohama}
\affiliation{Korea University, Seoul}
\affiliation{Kyoto University, Kyoto}
\affiliation{Kyungpook National University, Taegu}
\affiliation{Institut de Physique des Hautes \'Energies, Universit\'e de Lausanne, Lausanne}
\affiliation{University of Ljubljana, Ljubljana}
\affiliation{University of Maribor, Maribor}
\affiliation{University of Melbourne, Victoria}
\affiliation{Nagoya University, Nagoya}
\affiliation{Nara Women's University, Nara}
\affiliation{National Kaohsiung Normal University, Kaohsiung}
\affiliation{National Lien-Ho Institute of Technology, Miao Li}
\affiliation{Department of Physics, National Taiwan University, Taipei}
\affiliation{H. Niewodniczanski Institute of Nuclear Physics, Krakow}
\affiliation{Nihon Dental College, Niigata}
\affiliation{Niigata University, Niigata}
\affiliation{Osaka City University, Osaka}
\affiliation{Osaka University, Osaka}
\affiliation{Panjab University, Chandigarh}
\affiliation{Peking University, Beijing}
\affiliation{Princeton University, Princeton, New Jersey 08545}
\affiliation{RIKEN BNL Research Center, Upton, New York 11973}
\affiliation{Saga University, Saga}
\affiliation{University of Science and Technology of China, Hefei}
\affiliation{Seoul National University, Seoul}
\affiliation{Sungkyunkwan University, Suwon}
\affiliation{University of Sydney, Sydney NSW}
\affiliation{Tata Institute of Fundamental Research, Bombay}
\affiliation{Toho University, Funabashi}
\affiliation{Tohoku Gakuin University, Tagajo}
\affiliation{Tohoku University, Sendai}
\affiliation{Department of Physics, University of Tokyo, Tokyo}
\affiliation{Tokyo Institute of Technology, Tokyo}
\affiliation{Tokyo Metropolitan University, Tokyo}
\affiliation{Tokyo University of Agriculture and Technology, Tokyo}
\affiliation{Toyama National College of Maritime Technology, Toyama}
\affiliation{University of Tsukuba, Tsukuba}
\affiliation{Utkal University, Bhubaneswer}
\affiliation{Virginia Polytechnic Institute and State University, Blacksburg, Virginia 24061}
\affiliation{Yokkaichi University, Yokkaichi}
\affiliation{Yonsei University, Seoul}
  \author{Y.~Enari}\affiliation{Nagoya University, Nagoya} 
  \author{K.~Abe}\affiliation{High Energy Accelerator Research Organization (KEK), Tsukuba} 
  \author{K.~Abe}\affiliation{Tohoku Gakuin University, Tagajo} 
  \author{T.~Abe}\affiliation{High Energy Accelerator Research Organization (KEK), Tsukuba} 
  \author{I.~Adachi}\affiliation{High Energy Accelerator Research Organization (KEK), Tsukuba} 
  \author{H.~Aihara}\affiliation{Department of Physics, University of Tokyo, Tokyo} 
  \author{Y.~Asano}\affiliation{University of Tsukuba, Tsukuba} 
  \author{T.~Aso}\affiliation{Toyama National College of Maritime Technology, Toyama} 
  \author{V.~Aulchenko}\affiliation{Budker Institute of Nuclear Physics, Novosibirsk} 
  \author{T.~Aushev}\affiliation{Institute for Theoretical and Experimental Physics, Moscow} 
% ...
\collaboration{The Belle Collaboration}

\date{\today}

\begin{abstract}

We have searched for the Lepton Flavor Violating decay 
$\tau^-\rightarrow\mu^-\eta$ using 
a data sample of 84.3 fb$^{-1}$  accumulated with the Belle detector 
at KEK. The  $\eta$-meson was detected through the decay modes:
$\eta\rightarrow\gamma\gamma$ and $\pi^+\pi^-\pi^0$. 
No signal candidates are found, and we obtain an upper limit for
the branching fraction 
${\cal B}(\tau^-\rightarrow\mu^-\eta)<3.4\times 10^{-7}$ 
at the 90\% confidence level.

\end{abstract}


\maketitle

\tighten

{\renewcommand{\thefootnote}{\fnsymbol{footnote}}}
\setcounter{footnote}{0}

Among the possible Lepton Flavor Violating (LFV) decays of the $\tau$-lepton, 
$\TME$ is the process 
that provides the most stringent bound on Higgs-mediated LFV. 
Sher~\cite{Sher} has pointed out that a flavor non-diagonal 
lepton-lepton-Higgs 
Yukawa coupling could be induced if slepton mixing is large.
The $\mu$-$\tau$-Higgs vertex is particularly promising since mixing
between left-handed smuons and staus is large in many
supersymmetric models~\cite{models}.
This mechanism initially led various authors~\cite{Babu} 
to study the enhancement
of the LFV decay $\T3M$ in the minimal supersymmetric standard model (MSSM). 
However, Sher's results indicate that $\TME$ is enhanced
by a factor of 8.4 compared to $\T3M$, due mainly to a color
factor and the mass-squared dependent Higgs coupling at the Higss-s-sbar
vertex.
In some models with reasonable assumptions
about MSSM parameters~[1,3] the $\TME$
branching fraction is given by
\begin{equation}
{\cal B}(\tau^- \!\! \rightarrow\mu^{\!-}\eta)= 
0.84\times 10^{-6} \! \times \! \left 
(\frac{\rm{tan}\beta}{60}\right)^{\!\!6} \!\!
\left(\frac{100~\rm{GeV}}{m_A} \right)^{\!\!4}\!\!,
\label{eq1}
\end{equation}
where $m_A$ is the pseudoscalar Higgs mass and $\rm{tan}\beta$ is the ratio 
of the vacuum expectation values $(\langle H_u\rangle/\langle H_d\rangle)$. 
In such models, $\TME$ and $\T3M$
are particularly sensitive to LFV at large  $\rm{tan}\beta$.

Previous experimental studies of $\TME$ by 
ARGUS~\cite{argus} and CLEO~\cite{cleo} set 90\% confidence level 
upper limits on the branching fraction of $7.3\times 10^{-5}$ from 0.387 fb$^{-1}$ of data, and $9.6\times 10^{-6}$ from 4.68 fb$^{-1}$ 
of data, respectively. 
We present here a new search based on a
data sample of 84.3 fb$^{-1}$, equivalent to 76.9M $\TT$ pairs, 
collected at the $\Upsilon(4S)$ resonance
 with the Belle detector
 at the KEKB asymmetric $e^+e^-$ collider~\cite{KEKB}. A description of  
the detector can be found in Ref.~\cite{belle}. 

For Monte Carlo (MC) studies, the following programs have been used to generate background (BG) events:
 KORALB/TAUOLA~\cite{tauola} for $\TT$ processes, QQ~\cite{qq} for $B\bar{B}$ and continuum, 
BHLUMI~\cite{bhabha} for Bhabha, KKMC~\cite{kkmc} for $\mu\mu$ and 
AAFH~\cite{aafhb} for two-photon processes.
The $\TME$ decay is initially assumed to
have a uniform angular distribution in the $\tau$'s rest frame. 
The Belle detector response is simulated by a GEANT3~\cite{geant} 
based program. 
Most kinematical variables are evaluated in the laboratory frame, 
unless denoted by the superscript ``CM'' in which case they are evaluated 
in the center-of-mass frame.  
Two $\eta$ decay modes are considered in this analysis: 
$\EGG~({\cal B}=39.43\pm 0.26\%)$ and $\E3P~({\cal B}=22.6\pm 0.4\%)$~\cite{pdg}.

For $\EGG$, we search for events containing exactly two 
oppositely charged tracks and two or more photons, two of which form 
an $\eta$.
The events should be consistent with a $\TT$ event, in which one $\tau$ decays 
to $\mu\eta$ and the other $\tau$ decays to a charged particle other than 
a muon with any number of $\gamma$'s and neutrinos.

\begin{figure}
\centerline{
\epsfxsize=8.0cm \epsfbox{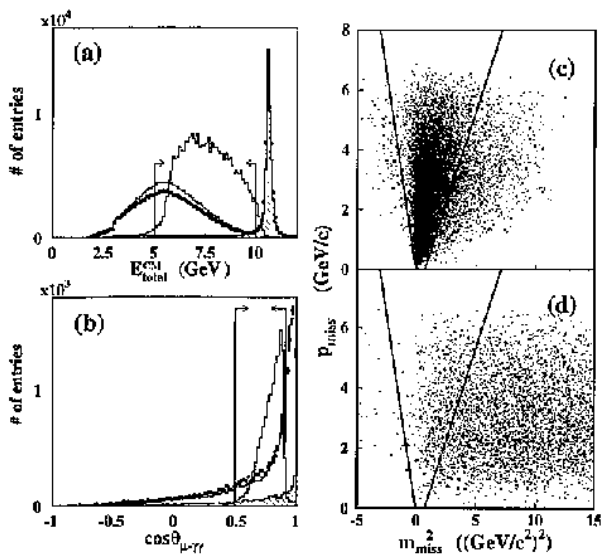} 
}
\caption{Some kinematical distributions from MC and data: 
  a) total energy, $E_{\rm total}^{CM}$, 
  b) cosine of the opening angle between the $\mu$ and $\gamma\gamma$ 
  on the signal side,
  c) $m_{\rm miss}^2$ vs. $p_{\rm miss}$ for signal MC events, 
  d) $m_{\rm miss}^2$ vs. $p_{\rm miss}$ for generic decays of $\TT$ MC. 
  The shaded histograms represent the signal $\TT$ MC, the hatched
  histograms represent the combined $\MM$ and Bhabha MC, the open
  histograms represent the combined $B\bar{B}$, continuum and
  two-photon MC and the closed circles represent the data.
  The selected regions in a), b) are indicated by the arrows.
      The selected region in c), d) is the area between the two lines: 
      $p_{miss}>-2.615 \times m_{miss}^2-0.191$  and 
      $p_{miss}>1.238 \times m_{miss}^2-0.869.$ 
} 
\label{level2-F}
\end{figure}

To select candidate events we require
the momentum of each track, $p$, and the energy of each
photon, $E_\gamma$, to satisfy $p > 0.1$ GeV/c and $E_\gamma > 0.1$
GeV.
The tracks and photons are required to be detected 
in the barrel or endcap of Belle: $-0.866<\cos\theta < 0.956$. 
To exclude Bhabha, $\mu\mu$ and two-photon events,
the total energy is constrained between 5 and 10 GeV in the CM frame, 
 as shown in Fig.~\ref{level2-F}(a).

In the CM frame the events are subdivided into two hemispheres by a plane 
perpendicular to the thrust axis. 
The signal side should contain a muon and two photons. 
The muon is identified as a track having 
a $\mu$ probability ${\cal{P}_{\mu}} > 0.9$~\cite{muid}.
An $\eta$ meson produced in a two-body $\tau$ decay has 
on average a higher momentum than $\eta$ mesons from other sources. 
Therefore, a photon from $\eta$ decay is 
required to have a rather high energy $E_{\gamma}>$ 0.22 GeV. 
To reduce background, events are rejected
when two $\gamma$'s, one from the signal side and the other from 
the tagging side $(\gamma')$, have a resolution-normalized 
$\pi^0$-mass in the range $-5<S^{\pi^0}_{\gamma\gamma'}<5$, 
 where $S^{\pi^0}_{\gamma\gamma'}=(m_{\gamma\gamma'}-0.135~\mbox{GeV/}c^2)/
\sigma^{\pi^0}_{\gamma\gamma'}$ and $\sigma^{\pi^0}_{\gamma\gamma'}$ 
is in the range 5-8 MeV$/c^2$. 
This $\pi^0$ veto rejects 86\% of BG events while retaining 75\% 
of the signal events. 
To further reduce BG events, the cosine of the opening angle between the
$\mu$ and $\gamma\gamma$ on the signal side is required to satisfy
$0.5 < \cos\theta_{\mu-\gamma\gamma} < 0.95$, as shown in Fig.~\ref{level2-F}(b).

In the tagging side hemisphere, the charged track should not be a muon 
(${\cal{P}}_{\mu}<0.6$ is imposed), but may be either an electron, 
 i.e. have an $e$ probability ${\cal{P}}_{e}>0.9$~\cite{muid} or 
a hadron (${\cal{P}}_{\mu}<0.6$ and ${\cal{P}}_{e}<0.9$). 
If an electron is found, the number of photons and the electron momentum are 
constrained by $n_{\gamma}\leq$ 2 and $p_e>$ 0.7 GeV/c.
If a hadron is found, the constraints are $n_{\gamma}\geq 0$ and 
$p_{\rm{had}}>$ 0.1 GeV/c.

 The following two criteria are imposed on the missing momentum 
and energy in the event. 
To ensure that the missing particles are neutrinos rather than 
$\gamma$'s or charged particles that 
fall outside of the detector acceptance, we require that the direction 
of the missing momentum should satisfy $-0.866< \cos\theta_{\rm{miss}}<0.956$.
Because neutrinos are emitted only on the tagging side, the direction of 
the missing momentum should be contained on the tagging side:
$\cos\theta_{\rm{thrust-miss}}<-0.55$. 
The correlation between the missing momentum, $p_{\rm{miss}}$, and 
the missing mass squared, $m_{\rm{miss}}^2$, shown in 
 Figs.~\ref{level2-F}(c) and (d) for 
 signal and generic $\TT$ MC, is utilized for additional BG rejection. 

\begin{figure}
\centerline{
\epsfxsize=9cm  \epsfbox{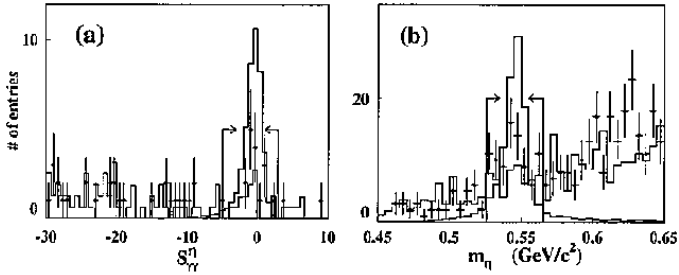} 
}
\caption{(a) invariant mass of $\gamma\gamma$ in terms of the resolution 
     normalized $\eta$-mass, $S^{\eta}_{\gamma\gamma}$, in the $\EGG$ case, and
     (b) $\eta$-mass from $\E3P$ reconstruction.  
     Signal and generic $\TT$ MC distributions are indicated by the shaded and open histograms, respectively.
     The selection region is indicated by the arrows.
}
\label{NewL2-F}
\end{figure}

The $\eta$ candidate is selected based on the signal-side $\gamma\gamma$ 
invariant mass in terms of 
the resolution-normalized $\eta$-mass, $-5 < S^{\eta}_{\gamma\gamma} < 3$, 
where $S^{\eta}_{\gamma\gamma}=(m_{\gamma\gamma}-0.547~\mbox{GeV/}c^2)/
\sigma^{\eta}_{\gamma\gamma}$ and $\sigma^{\eta}_{\GG}$
 is 12 MeV$/c^2$. 
The resulting $S^{\eta}_{\gamma\gamma}$ distributions for signal and generic 
$\TT$ MC and data are shown in Fig.~\ref{NewL2-F}(a).

The application of these selection criteria to the data set results 
in a total yield of 18 events.
The detection efficiency is measured from MC studies to be 
$\epsilon(2\gamma)=9.3\%$. 
In MC, small backgrounds from the three following processes survive: 
8.6$\pm$2.2 events from generic $\TT$, 2.5$\pm$1.8 events from 
$\mu\mu$ and 5.8$\pm$2.2 events from the continuum.

For the $\E3P$ mode, we search for events containing four charged tracks 
(net charge = 0) and two or more photons. 
Because of the higher multiplicity compared to the $\EGG$ mode 
the detection efficiency is smaller; however, the extra reconstruction 
constraint in the $\eta$ decay chain improves the background rejection power.
The selection criteria are similar to those in the $\EGG$ case with 
the differences listed below.

The minimum photon energy is reduced from 0.1 GeV to 0.05 GeV, 
since the photons from this decay mode have a softer 
energy distribution compared to those in $\EGG$. 
The signal side hemisphere should have three tracks and two or more photons.
One track must be a muon (${\cal{P}}_{\mu}>$ 0.9),
but particle identification is not performed on the other two tracks
--- they are treated as pions.
We also require that one $\pi^0$ be reconstructed from the photons in 
the signal hemisphere, such that $-5 < S_{\GG}^{\pi^0} < 5$.  
Figure~\ref{NewL2-F}(b) shows the reconstructed mass of $\eta$. 

After the cuts, 67 events remain in the data, while the generic $\TT$ MC
predicts a contribution of 38.0$\pm$4.6 events, 
and the continuum MC predicts 15.6 $\pm$ 3.5 events. 
The detection efficiency is $\epsilon(3\pi)=5.6\%$.

\begin{figure}
\begin{center}
\centerline{
\epsfxsize=6.1cm  \epsfbox{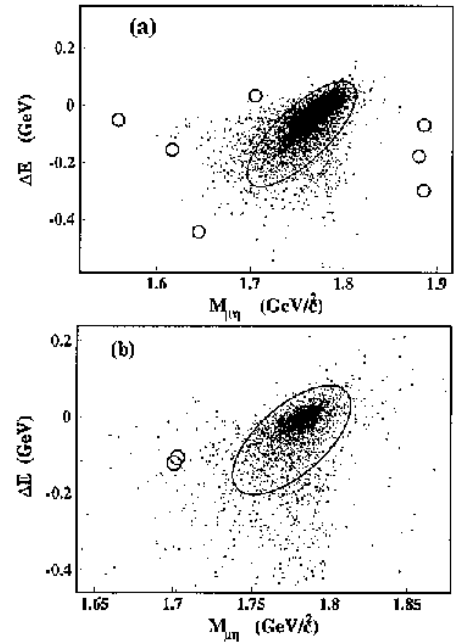}
}
\caption{
Final event distributions over a $\pm 10\sigma$ region 
   in the $\Minv$--$\DelE$ plane for (a) the $\EGG$ and (b) $\E3P$ modes. 
   The ellipses are the signal regions with an acceptance of 
   $\Omega=90\%$.
   The data are indicated by the open circles, and the signal MC events are 
plotted as dots.
}
\label{open}
\end{center}
\end{figure}

 The final evaluation of the number of signal candidates is performed 
by defining a signal-region in the $\Minv$--$\DelE$ plane,  
where the candidate $\ME$ system should have an invariant mass 
($\Minv$) close 
to the $\tau$-lepton mass and an energy close to the beam-energy in the CM 
frame, i.e. $\DelE=E^{\rm{CM}}_{\ME}-E^{\rm{CM}}_{\rm{beam}}\simeq 0$. 
Figures~\ref{open}(a) and \ref{open}(b) show  scatterplots of the 
signal MC in the
$\Minv$--$\DelE$ plane for the $\EGG$ and $\E3P$ modes, respectively. 
The signal exhibits a long low-energy tail 
due to initial-state radiation and calorimeter energy 
leakage for photons. 
By reproducing the $\Minv$ and $\DelE$ distributions around 
the peak with an asymmetric Gaussian function, the $\Minv$ and 
$\DelE$ resolutions are evaluated to be 
$\sigma_{\Minv}^{\rm low/high}$= 25.8$\pm$0.7/15.3$\pm$0.4 MeV/$c^2$ and 
$\sigma_{\DelE}^{\rm low/high}$= 69.7$\pm$3.0/34.7$\pm$1.2 MeV 
for the $\EGG$ mode, and
$\sigma_{\Minv}^{\rm low/high}$=  13.8$\pm$0.4/9.0$\pm$0.4 MeV/$c^2$ and
$\sigma_{\DelE}^{\rm low/high}$= 44.4$\pm$2.3/22.6$\pm$1.3 MeV for the
$\E3P$ mode, 
where the ``low/high'' superscript indicates the lower/higher energy side 
of the peak. 
To optimize the sensitivity, we take an elliptically shaped 
 signal region in the $\Minv$--$\DelE$ plane, with a signal 
acceptance of $\Omega$=90\%,  as shown in Fig.~\ref{open}. 


Figure \ref{open} shows the final data distributions for 
a $\pm 10\sigma_{\Minv/ \DelE}$ 
region in the $\Minv$--$\DelE$ plane. In the signal region,
there are no events in either the data or background MC. Outside
the signal region, 7 events for the $\EGG$ mode and 2 events
for the $\E3P$ mode are observed in data, while MC predicts
 $3.7\pm 2.4$ and $0.0^{+4.0}_{-0.0}$ events, respectively.
The observed data yields are consistent with MC.
The BG yield in the signal region, estimated
from the sidebands, is found to be
0.5$\pm$0.2 for $\EGG$ and 0.0$^{+0.4}_{-0.0}$ events for $\E3P$.

%

As no events are observed, an upper limit on the number
of events is set using a Bayesian approach, which gives $s_0=2.3$ at 90\% C.L.
The upper limit on the branching fraction, at 90\% C.L., is
given by
\begin{equation}
{\cal B}(\TME) < \frac{s_0}{2~(\epsilon\Omega\times {\cal B}_{\eta})
\times N_{\TT}},
\label{eq4}
\end{equation}
 where ${\cal B}_{\eta}$ is the branching fraction of $\eta$--decay to either $\gamma\gamma$ or
$\pi^+\pi^-\pi^0$.
The calculated upper limits, at 90\% C.L., are
thus found to be $4.6 \times 10^{-7}$ for the $\EGG$ mode, and 
$13.1 \times 10^{-7}$  for the $\E3P$ mode.
Combining the two decay modes, we obtain $\epsilon\Omega\times 
{\cal B}_{\eta}=4.4\%$ and ${\cal B}(\TME)~<~3.4\times 10^{-7}$ at  90\% C.L.

 The systematic uncertainties on the detection sensitivity,
$2(\epsilon\Omega\times {\cal B}_{\eta})\times N_{\TT}$, 
 arise from the track reconstruction efficiency (2.0\% in the $\EGG$ mode 
and 2.0\% in the $\E3P$ mode), $\eta$ reconstruction efficiency 
(2.0\% and 4.2\%, the latter of which includes the uncertainties 
of tracking efficiency for $\E3P$), $\pi^0$ veto (5.5\% and none),
muon identification efficiency (4.0\% and 4.0\%), trigger efficiency 
(1.4\% and 1.4\%), beam background (2.3\% and 2.1\%), luminosity 
(1.4\% and 1.4\%), ${\cal B}_{\eta}$ (0.7\% and 1.8\%) and MC statistics 
(1.3\% and 2.1\%).
Adding all of these components in quadrature, the total uncertainty is
evaluated to be 8.1\% for 
$\EGG$ and 7.3\% for $\E3P$. 
For the combination of the two decay modes the systematic 
uncertainty is  $\pm 7.9\%$.

This systematic uncertainty is included in the upper limit 
following Ref.~\cite{Cous}, where the detection sensitivity, 
$2(\epsilon\Omega \times {\cal B}_{\eta})N_{\tau\tau}$, is modelled by a
Gaussian distribution having a width given by the systematic error
quoted above.
There is no appreciable effect on the branching fraction, ${\cal B}$.

The angular distribution of the $\TME$ decay has a
strong dependence on the LFV interaction structure~\cite{spin}
 and spin correlations between the $\tau$'s
at the signal and tagged sides must be considered. 
To evaluate the maximum possible variation, V$-$A and V+A interactions
are assumed; no statistically significant difference 
in the $\Minv$--$\DelE$ distribution or in the efficiency
is found compared to the case of the uniform distribution.

 As a result, we obtain an upper limit on the branching fraction 
for the Lepton Flavor Violating $\TME$ decay of
\begin{equation}
{\cal B}(\TME)~<~3.4\times 10^{-7},
\end{equation}
at 90\% C.L. 
This result improves the previous upper limit, 
${\cal B}(\TME)<9.6\times 10^{-6}$~\cite{cleo}, by a factor of 30. 

\begin{figure}
\begin{center}
\epsfxsize=7.5cm  \epsfbox{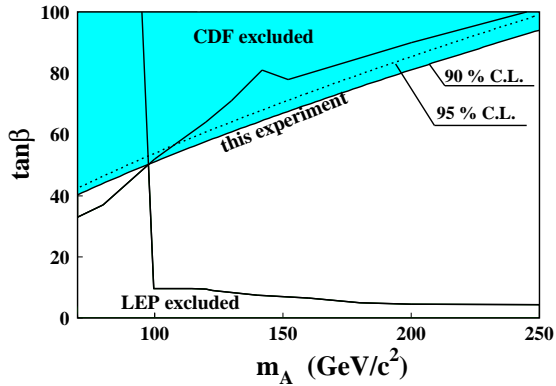} 
\caption{Experimentally excluded $m_A-\tan\beta$ parameter space. 
     The result of this experiment using \cite{Sher}
     is indicated by the shaded region 
     together with the regions excluded by LEP~\cite{LEP} 
     and the Tevatron~\cite{pdg,Tevatron}.} 
\label{ma-tanb}
\end{center}
\end{figure}

Using Eq.(\ref{eq1}), which was derived in a seesaw MSSM
with a specific neutrino mass texture, our upper limit
 restricts the allowed parameter space for 
$m_A$ and $\tan\beta$, as indicated in Fig.~\ref{ma-tanb}, where 
our boundary is indicated in the cases of 90\% and 95\% C.L.
Figure~\ref{ma-tanb} also shows the 95\% C.L. constraints from 
high energy collider experiments at LEP~\cite{LEP} and CDF~\cite{Tevatron}. 
Our result has a sensitivity close to 
that of the CDF experiment, achieved by searching for  
$pp\rightarrow A/\phi b\overline{b}\rightarrow b\overline{b}b\overline{b}$, 
where $\phi$ is a CP-even neutral Higgs state and $A$ is a CP-odd state 
in the MSSM.

We wish to thank the KEKB accelerator group for the excellent
operation of the KEKB accelerator.
We are grateful to Y.~Okada for fruitful discussions
on theoretical aspects of $\tau$ spin correlations.
We also thank A.~Dedes and J.~Hisano for important
and useful comments about model dependence of limits on
SUSY parameters.
We acknowledge support from the Ministry of Education,
Culture, Sports, Science, and Technology of Japan
and the Japan Society for the Promotion of Science;
the Australian Research Council
and the Australian Department of Industry, Science and Resources;
the National Science Foundation of China under contract No.~10175071;
the Department of Science and Technology of India;
the BK21 program of the Ministry of Education of Korea
and the CHEP SRC program of the Korea Science and Engineering
Foundation;
the Polish State Committee for Scientific Research
under contract No.~2P03B 01324;
the Ministry of Science and Technology of the Russian Federation;
the Ministry of Education, Science and Sport of the Republic of
Slovenia;
the National Science Council and the Ministry of Education of Taiwan;
and the U.S.\ Department of Energy.


\begin{thebibliography}{99}
\bibitem{Sher} M.~Sher, Phys. Rev. D~{\bf 66}, 057301 (2002).
\bibitem{models} 
I.~Hinchliffe and F.E.~Paige, Phys. Rev. D~{\bf 63}, 115006 (2001); 
J.~Hisano, T.~Moroi, K.~Tobe and M.~Yamaguchi, Phys. Rev. D~{\bf 53}, 
2442 (1996).
\bibitem{Babu} K.S.~Babu and C.~Kolda, Phys. Rev. Lett. {\bf 89}, 
241802 (2002);  
A.~Dedes, J.~Ellis and M.~Raidal, Phys. Lett. B~{\bf 549}, 159 (2002). 
The latter results in a branching fraction six times 
smaller than that predicted by Babu and Kolda. 
Eq.(\ref{eq1}) taken from \cite{Sher} is based on the result of Babu and Kolda.
\bibitem{argus} H.~Albrecht et al., ARGUS Collaboration, 
Z. Phys. C {\bf 55}, 179 (1992).
\bibitem{cleo} G.~Bonvicini et al., CLEO Collaboration, 
Phys. Rev. Lett. {\bf 79}, 1221 (1997).
\bibitem{KEKB} S.~Kurokawa and E.~Kikutani, 
Nucl. Instr. Meth. A~{\bf 499}, 1 (2003), and other papers 
included in this Volume.
\bibitem{belle} A.~Abashian et al., Belle Collaboration, 
Nucl. Instr. Meth. A~{\bf 479}, 117 (2002). 
\bibitem{tauola} S.~Jadach and Z.W\c{a}s, 
Comp. Phys. Commun. {\bf 85}, 453 (1995).
\bibitem{qq} http://www.lns.cornell.edu/public/CLEO/soft/QQ/.
\bibitem{bhabha} S.~Jadach, E.~Richter-W\c{a}s, B.F.L.~Ward and Z.~W\c{a}s, 
Comp. Phys. Commun. {\bf 70}, 305 (1992).
\bibitem{kkmc} S.~Jadach, B.H.L.~Ward and Z.~W\c{a}s, 
Comp. Phys. Commun. {\bf 130}, 260 (2000).
\bibitem{aafhb} F.A.~Berends, P.H.~Daverveldt and R.~Kleiss, 
Comp. Phys. Commun. {\bf 40}, 285 (1986).
\bibitem{geant} R.~Brun et al., GEANT 3.21 CERN Report No. DD/EE/84-1, 453.
\bibitem{pdg} K.~Hagiwara et al., Phys. Rev. D~{\bf 66}, 010001 (2002).
\bibitem{muid} A.~Abashian et al., Nucl. Instr. Meth. A~{\bf 491}, 69 (2002).
\bibitem{Cous} R.~Cousins and V.~Highland, 
Nucl. Instr. Meth. A~{\bf 320}, 331 (1992).
\bibitem{spin} R.~Kitano and Y.~Okada, Phys. Rev. D~{\bf63}, 113003 (2001).
\bibitem{LEP} {LEP~Higgs~Working~Group}, 
http://lephiggs.web.cern/ LEPHIGGS/papers/ Note 2001-04.
\bibitem{Tevatron} T.~Affolder et al., CDF Collaboration, 
Phys. Rev. Lett. {\bf 86}, 4472 (2001).

\end{thebibliography}
\end{document}